\documentclass[aps,prb,twocolumn,amsmath,showpacs,superscriptaddress,amssymb]{revtex4}

\usepackage{graphics}
\usepackage{amsmath}
\usepackage{dcolumn}
\usepackage{citesort}

\begin{document}

\title {Non-Adiabatic Potential-Energy Surfaces\\ by Constrained Density-Functional Theory}

\author{J\"org Behler}
\affiliation{Fritz-Haber-Institut der Max-Planck-Gesellschaft,
Faradayweg 4-6, D-14195 Berlin, Germany}
\author{Bernard Delley}
\affiliation{Paul-Scherrer-Institut,
WHGA/123, CH-5232 Villigen PSI, Switzerland}
\author{Karsten Reuter}
\affiliation{Fritz-Haber-Institut der Max-Planck-Gesellschaft,
Faradayweg 4-6, D-14195 Berlin, Germany}
\author{Matthias Scheffler}
\affiliation{Fritz-Haber-Institut der Max-Planck-Gesellschaft,
Faradayweg 4-6, D-14195 Berlin, Germany}

\date{\today}

\begin{abstract}
Non-adiabatic effects play an important role in many chemical
processes. In order to study the underlying non-adiabatic
potential-energy surfaces (PESs), we present a locally-constrained
density-functional theory approach, which enables us to confine
electrons to sub-spaces of the Hilbert space, e.g. to selected atoms
or groups of atoms. This allows to calculate non-adiabatic PESs for
defined charge and spin states of the chosen subsystems. The
capability of the method is demonstrated by calculating
non-adiabatic PESs for the scattering of a sodium and a chlorine
atom, for the interaction of a chlorine molecule with a small metal
cluster, and for the dissociation of an oxygen molecule at the
Al(111) surface.
\end{abstract}


\maketitle

\section{Introduction}

Assuming that electrons can react much faster to external
perturbations than nuclei, the Born-Oppenheimer
approximation~\cite{born27} (BOA) is a widely employed approach to
separate electronic and nuclear motion in dynamic processes. The
nuclei then appear static to the electrons, which in turn set up a
potential governing the motion of the nuclei. Formally the approach
starts with the general many-body Schr\"odinger equation
\begin{equation}
H \Psi \;=\; \left( T^{\rm nuc} + V^{\rm nuc-nuc} + H^e \right) \Psi \;=\; E \Psi \quad ,
\label{generalH}
\end{equation}
where $T^{\rm nuc}$ is the kinetic energy operator of the nuclei,
$V^{\rm nuc-nuc}$ the interaction potential of the nuclei, and $H^e$
the electronic Hamiltonian containing the electron kinetic energy,
as well as the electron-electron and electron-nuclei interactions.
In an adiabatic representation \cite{omalley71,child74,darling95},
the general dependence of the full many-body wavefunction $\Psi(\{
\mathbf{r}_i \sigma_i\},\{ \mathbf{R}_I \sigma_I \})$ on the
position and spin coordinates of all electrons, $\mathbf{r}_i$ and
$\sigma_i$, and on the position and spin coordinates of all nuclei,
$\mathbf{R}_I$ and $\sigma_I$, is written as
\begin{equation}
\Psi \;=\; \sum_{\nu} \; \Lambda^{\rm adia}_{\nu}(\{ \mathbf{R}_I \sigma_I \}) \; \Phi^{\rm adia}_{\nu}(\{ \mathbf{r}_i \sigma_i \}, \{ \mathbf{R}_I \sigma_I \}) \quad ,
\label{adiabatic}
\end{equation}
where the $\Phi^{\rm adia}_{\nu}$ are chosen to be the
eigenfunctions of the electronic Hamiltonian at the actual position
of the nuclei
\begin{equation}
H^e \Phi^{\rm adia}_{\nu} \;=\; E^e_{\nu} \Phi^{\rm adia}_{\nu}  \quad .
\label{electronicH}
\end{equation}
Inserting Eq. (\ref{adiabatic}) into Eq. (\ref{generalH}) leads to a
set of equations for the wavefunctions of the nuclei
\begin{eqnarray}
E \Lambda_{\nu}^{\rm adia} = \left( T^{\rm nuc} + V^{\rm nuc-nuc} + E^e_{\nu} \right) \Lambda^{\rm adia}_{\nu} \nonumber \\
+ \mbox{non-adiabaticity terms} \;\; ,
\label{adiamotion}
\end{eqnarray}
in which the ``non-adiabaticity terms'' summarize matrix elements of
the momentum and kinetic energy operators of the nuclear motion. The
Born-Oppenheimer approximation corresponds to setting these terms to
zero, which implies that the electrons assume their electronic state
instantaneously for any position of the nuclei, unaffected by the
nuclear dynamics. The nuclei are then moving on the potential-energy
surface (PES) $V^{\rm adia}_{\nu} = V^{\rm nuc-nuc} + E^e_{\nu}$,
also called the Born-Oppenheimer surface of the electronic state
$\nu$.

If, in the BOA, the system is initially in the electronic ground
state, it will remain there irrespective of the dynamics of the
nuclei. However, in real life electrons may not be able to follow
the motion of the nuclei instantaneously, and, e.g. when selection
rules apply, they may find themselves in an excited state. For
example, chemical reactions forming singlet molecules from triplet
and singlet reactants are forbidden by Wigner's spin selection rule.
And the triplet multiplicity is the actual reason, why most
reactions of O$_2$ with other molecules or substances, although
being exothermic, do not proceed at room temperature; they are
kinetically hindered. In other words, in a chemical reaction the
spin of the reactants must be conserved or transferred to some other
entity. And the transition from the O$_2$ ground state,
($^3\Sigma_g^-$) to the first excited state ($^1\Delta_g$), is
strictly forbidden for the isolated molecule, as is the reverse
(de-excitation) process, once the molecule has been excited to the
$^1\Delta_g$ state. Thus, when probabilities for transitions between
different electronic states are low, e.g. due to selection rules,
the assumption that the system will always remain in the electronic
ground state may become incorrect. For the mentioned O$_2$ example
this implies that when an external field shifts the $^1\Delta_g$
energy below the $^3\Sigma_g^-$ energy, the probability for an
electronic transition toward the $^1\Delta_g$ state will be low.
Indeed, this is an important aspect of the O$_2$/Al(111)
interaction, one of our examples discussed below.

For such, or alike, situations it is necessary to go beyond the BOA
and to consider the coupling between different
states.\cite{zener32,lichten63} For a full description the
non-adiabaticity terms must be calculated, which in practice means
to evaluate the matrix elements for the derivative couplings of the
nuclear and electronic motion. For this, the use of another set of
electronic functions $|\Phi^{\rm dia}_{\mu}>$ in the wave function
expansion in Eq. (\ref{adiabatic}) may be suitable. The idea behind
such ``diabatic'' states is that in a dynamic (scattering) event,
the electrons lag behind the nuclear motion and thus tend to
conserve a given (initial) character of the electronic structure.
Formally, the $|\Phi^{\rm dia}_{\mu}>$ are then constructed to
minimize the nuclear derivative coupling terms, such that the matrix
representation of the nuclear momentum operator becomes diagonal in
the diabatic basis \cite{lichten63,smith69,pacher93}. The equivalent
to Eq. (\ref{adiamotion}) in a diabatic representation is thus a
matrix equation
\begin{eqnarray}
\sum_{\nu} \left( T^{\rm nuc} \delta_{\mu\nu} + V^{\rm nuc-nuc} \delta_{\mu\nu} + H^e_{\mu\nu} \right)  \Lambda^{\rm dia}_{\nu} \nonumber \\
= E_\mu  \Lambda^{\rm dia}_{\mu} \quad ,
\label{diamotion}
\end{eqnarray}
where the diagonal elements of the potential, $V_{\mu}^{\rm dia} =
V^{\rm nuc-nuc} + H^e_{\mu\mu}$, are the diabatic PESs, and the
off-diagonal elements, $H^e_{\mu\nu}$, describe the coupling between
PES $\mu$ and PES $\nu$. For the above example of an O$_2$ molecule,
a suitable diabatic state would e.g. be given by the
${}^3\Sigma_g^-$ state of the isolated molecule, and the
corresponding diabatic PES can be used to describe the motion of a
molecule that tries to conserve this electronic configuration, even
in regions where external fields (or interactions with other
species) bring the ${}^1\Sigma_g$ below the ${}^3\Sigma_g^-$ energy.

The aim of this work is to develop and apply a non-adiabatic
approach, with a similar physical motivation as the diabaticity
concept. We account for the tendency to maintain a given character
of the electronic structure, e.g. due to selection rules, by
focusing on the dynamics of a system with defined charge and spin in
suitably chosen subsystems, typically the two reactants in a
scattering event. Using again the example of an O$_2$ molecule, this
could e.g. be a state with fixed triplet spin on the molecule, even
when it interacts with another reactant. Instead of using properties
of the many-body wavefunction as formal basis, we thus build our
approach on a defined total spin (or total charge, or other well
defined quantity) in the local Hilbert space of a chosen subsystem,
which can be an atom or a group of atoms. The thus defined
non-adiabatic PESs are computed with density-functional theory (DFT)
\cite{parr89,dreizler90}, and in particular using the idea of
constrained DFT formulated by Dederichs {\em et al.} in 1984
\cite{dederichs84}. The basic idea is to modify the energy
functional employed in DFT by applying physically meaningful
constraints. The constraint is enforced through an additional
Lagrange multiplier, which on the level of the Kohn-Sham (KS)
equations yields an additional constraint term to the effective KS
potential. In the present work we employ a local constraint via a
projection technique that enables us to freeze the charge and spin
states of the chosen subsystems.

Having stated this general philosophy, two points deserve a critical
discussion. First, we point out that in practice our non-adiabatic
states may often be quite close to those of the diabaticity concept.
Still, the formal definition is different, and in some cases like
the example of the O$_2$/Al(111) system discussed below, there are
notable differences. In contrast to recent other works in this area
\cite{prezhdo99,wu06}, we therefore prefer to refer to our states
simply as ``non-adiabatic states'' to underscore this difference to
the established diabaticity concept. As a second point, we note that
a solid, mathematical proof of the validity of constrained DFT does
not exist. However, spin-density functional theory, the Slater-Janak
transition state concept \cite{dederichs84,gunnarsson76,slater28},
the fixed-spin moment (FSM) approach \cite{schwarz84} and some other
examples present frequently employed, important, and successful
applications of the method. Our concept represents a mild
generalization of such applications, and our constraint is plausible
and physically meaningful: The spins or charges of two interacting
systems may be hindered to adjust or combine when there are
selection rules. Thus, the appropriate total energy surfaces of the
scattering event should (up to a certain distance) be that of
conserved local spins or charges.

In a previous publication we have already used the present approach
to resolve a long-standing problem in surface science, namely the
low sticking probability of oxygen molecules at the Al(111)
surface.\cite{behler05} However, the method is much more general, as
we will illustrate below by also applying it to two other,
non-periodic model systems, namely a scattering event of a sodium
and a chlorine atom, and the interaction of a Cl$_2$ molecule with a
magnesium cluster. In this respect, we also mention notable,
independent work of Wu and Van Voorhis~\cite{wu06,wu05}, which is
essentially equivalent to our approach~\cite{behler05} in that also
in their work an additional potential is introduced to constrain
electron numbers in well-defined parts of a system. As in our
approach this potential is determined in a self-consistent way using
formally identical methodology, and has been employed to study
charge-transfer reactions.

\section{Locally-Constrained Density-Functional Theory}

Our locally-constrained DFT (LC-DFT) approach starts by assigning
electrons to defined sub-spaces of the total Hilbert space, e.g. to
atoms or groups of atoms. This is done by employing a suitable
projection scheme to distinguish the individual subsystems. In the
following sections we will present this formalism for a system
consisting of two subsystems called A and B, with straightforward
generalization to more than two subsystems. Considering electrons
and their spin, this leads to four electron numbers $N_{\rm
A}^{\uparrow}$, $N_{\rm A}^{\downarrow}$, $N_{\rm B}^{\uparrow}$ and
$N_{\rm B}^{\downarrow}$, which uniquely define the PES $V^{\rm
na}_{N_{\rm A}^{\uparrow}, N_{\rm A}^{\downarrow}, N_{\rm
B}^{\uparrow}, N_{\rm B}^{\downarrow}}(\{ \mathbf{R}_{I} \sigma_{I}
\})$ for the non-adiabatic quantum state with fixed spins and
charges of the subsystems. Expressing the constraints in terms of
auxiliary potentials, the electronic structure problem is solved
self-consistently using DFT, i.e., the electronic structure is fully
relaxed under the given constraints yielding $V^{\rm na}_{N_{\rm
A}^{\uparrow}, N_{\rm A}^{\downarrow}, N_{\rm B}^{\uparrow}, N_{\rm
B}^{\downarrow}}$.

\subsection{Definition of the Subsystems}

In order to assign electrons to the two subsystems, the Kohn-Sham
single-particle wavefunctions are expanded into localized,
atom-centered basis functions, e.g. Gaussians, Slater type orbitals,
or numerical orbitals. The implementation is therefore particularly
convenient in codes employing such basis sets, whereas in codes
based on other basis sets, like plane waves, the implementation
requires intermediate projection steps onto localized functions
\cite{scheffler84,sanchez95,scheffler00}. The Kohn-Sham orbital
$\phi_{i}^{\sigma}$ with index $i$ and spin index $\sigma
=\,\uparrow, \downarrow$ is thus written as a linear combination of
all basis functions $\chi_{j}$, or in calculations using periodic
boundary conditions as a linear combination of
$\mathbf{k}$-dependent Bloch basis functions $\chi_{j}^{\mathbf{k}}
= e^{i\mathbf{kr}}\chi_{j}$,
\begin{equation}
\phi_{i}^{\mathbf{k}\sigma} \;=\; \sum_{j=1}^{n}c_{ij}^{\mathbf{k}\sigma} \chi_{j}^{\mathbf{k}} \quad .
\label{spstate}
\end{equation}
In the following our derivation will refer to the periodic case,
while for finite systems the dependence on the $\mathbf{k}$-point
index would be simply dropped.

In the case of atom-centered basis functions each basis function is
uniquely assigned either to subsystem A or to subsystem B: All basis
functions centered at atoms being part of subsystem A define the
Hilbert space of subsystem A, and all basis functions centered at
atoms being part of subsystem B define the Hilbert space of
subsystem B. Every single-particle wavefunction can then be
projected onto the two Hilbert sub-spaces, which is done separately
for each $\mathbf{k}$-point and each spin, and taking into account a
non-orthogonality of the atomic orbitals by including the overlap
matrix elements $S_{jk}^{\mathbf{k}} = <\chi_{j}^{\mathbf{k}} |
\chi_{k}^{\mathbf{k}}> $,

\begin{subequations}
\label{parts}
\begin{eqnarray}
p_{{\rm A},i}^{\mathbf{k}\sigma} &=&
<\phi_i^{\mathbf{k}\sigma}|\phi_{i,\rm A}^{\mathbf{k}\sigma}>\;=\; \sum_{j=1}^n\sum_{k=1}^m c_{ij}^{\mathbf{k}\sigma}S_{jk}^{\mathbf{k}}c_{ik}^{\mathbf{k}\sigma} \label{pup} \label{partO2}\\
p_{{\rm B},i}^{\mathbf{k}\sigma} &=&
<\phi_i^{\mathbf{k}\sigma}|\phi_{i,\rm B}^{\mathbf{k}\sigma}>
\;=\; \sum_{j=1}^n\sum_{k=m+1}^n
c_{ij}^{\mathbf{k}\sigma}S_{jk}^{\mathbf{k}}c_{ik}^{\mathbf{k}\sigma} \; . \label{pdn} \label{partB}
\end{eqnarray}
\end{subequations}
Here, $p_{{\rm A},i}^{\mathbf{k}\sigma}$ is the projection onto
subsystem A, $p_{{\rm B},i}^{\mathbf{k}\sigma}$ the projection onto
subsystem B, and $n$ is the total number of basis functions, of
which the first $k=1\ldots m$ are the basis functions of subsystem
A. $\phi_{i,\rm A}^{\mathbf{k}\sigma}$ is defined as the A-component
of $\phi_i^{\mathbf{k}\sigma}$, i.e., all coefficients referring to
basis functions of subsystem B are set to zero. Correspondingly, all
coefficients referring to basis functions of subsystem A are set to
zero in the functions $\phi_{i,\rm B}^{\mathbf{k}\sigma}$ with
$i=m+1 \ldots n$, which define the complementary B-component of
$\phi_i^{\mathbf{k}\sigma}$. We note that the normalization
condition $p_{A,i}^{\mathbf{k}\sigma} + p_{B,i}^{\mathbf{k}\sigma} =
1$ holds for each state $i$ by construction, because each
$\phi_{i,\rm A}^{\mathbf{k}\sigma}+\phi_{i,\rm B}^{\mathbf{k}\sigma}
= \phi_i^{\mathbf{k}\sigma}$ is a normalized eigenstate. This
relation can be used to reduce the computational effort in that just
one of the two double sums in Eq.~(\ref{parts}) needs to be
calculated and the other is obtained from the normalization
condition. We note that in this scheme like in all projection
schemes the actual values of $p_{{\rm A},i}^{\mathbf{k}\sigma}$ and
$p_{{\rm B},i}^{\mathbf{k}\sigma}$ depend on the choice of the basis
sets defining the Hilbert spaces of the subsystems.

Having split each single-particle state into an A-part and a B-part
allows to construct the partial densities of states (pDOSs) for
subsystems A and B and for the two spin channels. Summing up the
resulting four pDOSs over all occupied single-particle states $i$
yields then the four electron numbers
\begin{subequations}
\begin{eqnarray}
N_{\rm A}^{\sigma} &=& \sum_{\mathbf{k}} \sum_{i} \sum_{j=1}^n \sum_{k=1}^m
f_i^{\mathbf{k}\sigma} c_{ij}^{\mathbf{k}\sigma} S_{jk}^{\mathbf{k}}
c_{ik}^{\mathbf{k}\sigma} \\
N_{\rm B}^{\sigma} &=& \sum_{\mathbf{k}} \sum_{i} \sum_{j=1}^n \sum_{k=m+1}^n
f_i^{\mathbf{k}\sigma} c_{ij}^{\mathbf{k}\sigma} S_{jk}^{\mathbf{k}} c_{ik}^{\mathbf{k}\sigma} \quad ,
\end{eqnarray}
\end{subequations}
where $f_i^{\mathbf{k}\sigma}$ is the occupation number of the
single-particle Kohn-Sham state $i$, typically chosen to be a Fermi
function. With this, also the total spin $\mathcal{S}$ of each of
the two subsystems is determined through  the difference of the
corresponding electron numbers
\begin{subequations}
\begin{eqnarray}
\mathcal{S}_{\rm A} &=& |N_{\rm A}^{\uparrow}-N_{\rm A}^{\downarrow}| \\
\mathcal{S}_{\rm B} &=& |N_{\rm B}^{\uparrow}-N_{\rm B}^{\downarrow}|
\quad .
\end{eqnarray}
\end{subequations}

\subsection{Constraining the Electron Numbers}

Having established the various electron numbers, we proceed to
introduce the constraint. Aiming to compute the non-adiabatic PES
$V^{\rm na}_{N_{\rm A}^{\uparrow}, N_{\rm A}^{\downarrow}, N_{\rm
B}^{\uparrow}, N_{\rm B}^{\downarrow}}(\{ \mathbf{R}_{I} \sigma_{I}
\})$ representing a defined spin and charge state in subsystems A
and B, we first distribute the corresponding $N_{\rm A}^{\uparrow}$,
$N_{\rm A}^{\downarrow}$, $N_{\rm B}^{\uparrow}$, and $N_{\rm
B}^{\downarrow}$ electrons into the four pDOSs derived from the set
of Kohn-Sham single-particle wavefunctions. This defines four Fermi
energies $\epsilon^{\uparrow}_{\rm F,A}$,
$\epsilon^{\downarrow}_{\rm F,A}$, $\epsilon^{\uparrow}_{\rm F,B}$
and $\epsilon^{\downarrow}_{\rm F,B}$, as well as four partial
electron densities, which sum to the total electron density. If the
Fermi energies were all degenerate at this stage, the chosen
non-adiabatic charge and spin state of the system would correspond
to the adiabatic ground state. However, typically the four Fermi
energies are all different, reflecting the fact that there is no
self-consistency between the total electron density and the
effective Kohn-Sham potential.

In order to achieve this self-consistency, we choose to align the
Fermi energies, still requiring that the electron numbers that can
be filled into the four pDOSs up to the resulting common Fermi level
remain unchanged. We therefore employ a method that shifts each of
the four pDOSs independently. This is particularly important when
hybridization between the two subsystems is present and a
single-particle state $i$ contains non-zero expansion coefficients
$c_{ij}^{\mathbf{k}\sigma}$ for basis functions belonging to
subsystem A, as well as for basis functions belonging to subsystem
B. In such cases, simply shifting the entire state $i$ as is e.g.
done in the $\Delta$SCF method would not change the relative
positions of the A and B Fermi energies. Instead, it is necessary to
act separately on the basis functions in both subsystems. Only this
gives the electronic structure enough flexibility to fully relax
under the imposed constrained electron numbers, which in the end
will yield a different expansion of state $i$ in terms of A and B
basis functions.

\begin{figure}[t!]
\scalebox{0.45}{\includegraphics{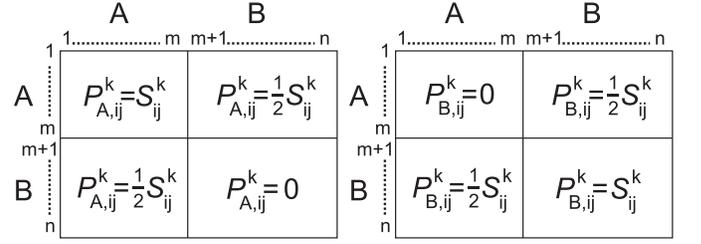}} \caption{Matrix
representation of the projection operators $\mathbf{P}_{\rm
A}^{\mathbf{k}}$ and $\mathbf{P}_{\rm B}^{\mathbf{k}}$ in the
Hilbert space of the Bloch basis functions $\chi_{j}^{\mathbf{k}}$.
The $S_{ij}^{\mathbf{k}}$ are the overlap matrix elements between
basis functions $i$ and $j$. $n$ is the total number of basis
functions and $m$ the number of basis functions of subsystem A.}
\label{fig1}
\end{figure}

In practice, we first align the Fermi energies separately for each
spin. Specifically, we request that $\epsilon^{\sigma}_{\rm F,A} =
\epsilon^{\sigma}_{\rm F,B}=\epsilon^{\sigma}_{\rm F}$, and
appropriately shift the A and B Fermi energies to the Fermi energy
$\epsilon^{\sigma}_{\rm F}$. This is achieved by adding to the
Kohn-Sham Hamiltonian auxiliary potentials that act differently on
the A and B basis functions. This auxiliary potentials consist of a
``strength'' factor $\Gamma^{\sigma}$ and a projection operator
$\mathbf{P}^{\mathbf{k}}$ onto the sub-spaces
\begin{subequations}\label{projector}
\begin{eqnarray}
\Gamma^{\sigma}_{\rm A} \mathbf{P}_{\rm A}^{\mathbf{k}} &=&
\frac{1}{2} \Gamma^{\sigma}_{\rm A} \cdot \big( \sum_{i=1}^m
|\chi_{i}^{\mathbf{k}}>< \chi^{i\mathbf{k}}|\nonumber \\&+&
\sum_{i=1}^m
|\chi^{i\mathbf{k}}>< \chi_{i}^{\mathbf{k}}| \big) \\
\Gamma^{\sigma}_{\rm B} \mathbf{P}_{\rm B}^{\mathbf{k}} &=&
\frac{1}{2} \Gamma^{\sigma}_{\rm B} \cdot \big( \sum_{i=m+1}^n
|\chi_{i}^{\mathbf{k}}>< \chi^{i\mathbf{k}}| \nonumber \\&+&
\sum_{i=m+1}^n |\chi^{i\mathbf{k}}>< \chi_{i}^{\mathbf{k}}| \big)
\quad ,
\end{eqnarray}
\end{subequations}
where the summation is done over the $m$ basis functions spanning
the Hilbert space of subsystem A and the $n-m$ basis functions
spanning the Hilbert space of subsystem B. In the chosen form
$\mathbf{P}^{\mathbf{k}}$ symmetrically contains covariant and
contravariant basis functions \cite{headgordon98,artacho91},
$\chi_i^{\mathbf{k}}$ and $\chi^{i,\mathbf{k}}$, respectively. The
resulting matrix representations of $\mathbf{P}^{\mathbf{k}}_{\rm
A}$ and $\mathbf{P}^{\mathbf{k}}_{\rm B}$ in the Hilbert space
spanned by the Bloch basis functions is then hermitian, which
facilitates the implementation into existing DFT codes as further
described below. As derived in the Appendix for subsystem A, the
form of these matrices is as shown schematically in Fig. \ref{fig1}:
The matrix element $P_{{\rm A},ij}^{\mathbf{k}}$ is equal to the
overlap matrix element $S_{ij}^{\mathbf{k}}$, if $i$ and $j$ both
refer to basis functions assigned to subsystem A. If only $i$ or $j$
refer to a basis function assigned to subsystem A, the matrix
element is $\frac{1}{2}\cdot S_{ij}^{\mathbf{k}}$. Finally, if
neither $i$ nor $j$ belong to subsystem A, $P_{{\rm
A},ij}^{\mathbf{k}}$ is zero, reflecting that the pDOS of subsystem
B is not affected by the auxiliary potential $\Gamma^{\sigma}_{\rm
A} \mathbf{P}_{\rm A}^{\mathbf{k}}$. The auxiliary potential
$\Gamma^{\sigma}_{\rm B} \mathbf{P}_{\rm B}^{\mathbf{k}}$ has an
analogous structure.

Since the purpose of the auxiliary potentials is to align the Fermi
energies of subsystems A and B with $\epsilon_{\rm F}^{\sigma}$, the
obvious choice for the ``strength'' factors $\Gamma^{\sigma}_{\rm
A}$ and $\Gamma^{\sigma}_{\rm B}$ is
\begin{subequations}\label{gamma}
\begin{eqnarray}
\Gamma^{\sigma}_{\rm A} &=& \epsilon^{\sigma}_{\rm F,A} -
\epsilon^{\sigma}_{\rm F} \\
\Gamma^{\sigma}_{\rm B} &=& \epsilon^{\sigma}_{\rm F,B} -
\epsilon^{\sigma}_{\rm F} \quad .
\end{eqnarray}
\end{subequations}
However, because of the resulting non-zero auxiliary potentials, the
initial single-particle states are no longer solutions of the new
effective Hamiltonian for each spin, comprised of both the Kohn-Sham
Hamiltonian {\em and} the auxiliary potential. As a result,
$\Gamma^{\sigma}_{\rm A}$ and $\Gamma^{\sigma}_{\rm B}$ must be
determined self-consistently. Diagonalization of the new effective
Hamiltonians yields new eigenvectors to construct new partial
densities of states, the Fermi levels of which define new strength
factors through Eq. (\ref{gamma}). This is repeated in a
self-consistency (sc) cycle, until the Fermi energies of subsystems
A and B are aligned to an arbitrary precision (for each spin
channel),
\begin{subequations}
\begin{eqnarray}
\epsilon_{\rm F}^{\uparrow} &=&
\epsilon_{\rm F,A}^{\uparrow} \;=\; \epsilon_{\rm F,B}^{\uparrow} \\
\epsilon_{\rm F}^{\downarrow} &=&
\epsilon_{\rm F,A}^{\downarrow} \;=\;
\epsilon_{\rm F,B}^{\downarrow} \quad .
\end{eqnarray}
\end{subequations}

The ensuing step of aligning the two different spin Fermi energies
$\epsilon_{\rm F}^{\uparrow}$ and $\epsilon_{\rm F}^{\downarrow}$ is
done in an analogous way, i.e. by adding another auxiliary potential
of the form of Eq. (\ref{projector}). In this case, the matrix
structure of the corresponding projection operator
$\mathbf{P}^{\mathbf{k}}$ onto one spin sub-space is simpler though.
Since this sub-space is spanned by all $n$ basis functions,
regardless of whether they are in subsystem A or B, the sum in Eq.
(\ref{projector}) goes up to $n$, and the matrix representation of
$\mathbf{P}^{\mathbf{k}}$ becomes simply the overlap matrix, cf.
Fig. \ref{fig1} with $m = n$. Adding an auxiliary potential
$\Gamma^{\sigma} \mathbf{P}^{\mathbf{k}}$ to the effective
Hamiltonian resulting from the preceding alignment of the A and B
Fermi energies, corresponds therefore to a mere shift of the
eigenvalues, depending on the chosen ``strength'' factor
$\Gamma^{\sigma}$. Here we choose to shift the spin-up and spin-down
pDOSs in opposite directions using $\Delta \epsilon_{\rm
F}=\frac{1}{N}(\epsilon_{\rm F}^{\uparrow}-\epsilon_{\rm
F}^{\downarrow})$,
$\Gamma^{\uparrow}=+N^{\downarrow}\cdot\Delta{\epsilon_{\rm F}}$ and
$\Gamma^{\downarrow}=-N^{\uparrow}\cdot\Delta{\epsilon_{\rm F}}$
with $N^{\uparrow}=N^{\uparrow}_{\rm A}+N^{\uparrow}_{\rm B}$,
$N^{\downarrow}=N^{\downarrow}_{\rm A}+N^{\downarrow}_{\rm B}$ and
$N=N^{\uparrow}+N^{\downarrow}$. As before, this procedure has to be
done in a self-consistent way, since adding the new auxiliary
potential modifies the effective Hamiltonian. In fact, since the
alignment of the A and B Fermi levels and the alignment of the
spin-up and spin-down Fermi levels is not independent of each other,
the two sc cycles must be nested. Discussing below how this can be
implemented in a numerically efficient way into existing DFT codes,
we note that once the double self-consistency is achieved, we arrive
at a final common Fermi level in the system and a new set of
single-particle states $i$ (with eigenvalues
$\epsilon_i^{\prime\sigma}$ and eigenvectors
$\phi_{i}^{\prime\sigma}$) that is the self-consistent solution to
the effective Hamiltonian, containing the original Kohn-Sham
Hamiltonian and the two auxiliary potentials. In each sub-space
spanned by the Bloch basis functions at one {\bf k}-point we
therefore have
\begin{eqnarray}
\nonumber H^{\bf k}_{\rm eff} \phi_{i}^{\prime\mathbf{k}\sigma} &=&
\left[ H_{\rm KS}^{\mathbf{k}} + \Gamma^{\sigma}_{\rm A,scf}
\mathbf{P}_{\rm A}^{\mathbf{k}} + \Gamma^{\sigma}_{\rm B,scf}
\mathbf{P}_{\rm B}^{\mathbf{k}}+ \Gamma^{\sigma}_{\rm scf}
\mathbf{P}^{\mathbf{k}}
\right] \phi_{i}^{\prime\mathbf{k}\sigma}  \\
&=& \epsilon_i^{\prime\mathbf{k}\sigma}
\phi_{i}^{\prime\mathbf{k}\sigma} \quad , \label{effhamilt}
\end{eqnarray}
with ``strength'' factor values $\Gamma^{\sigma}_{\rm A,scf}$,
$\Gamma^{\sigma}_{\rm B,scf}$ and $\Gamma^{\sigma}_{\rm scf}$ as
determined in the last cycle of the nested sc loops.

\subsection{Numerical Considerations}

At this stage it is appropriate to recall what has been achieved so
far. The imposed constraint of fixed electron numbers $N_{\rm
A}^{\uparrow}$, $N_{\rm A}^{\downarrow}$, $N_{\rm B}^{\uparrow}$,
and $N_{\rm B}^{\downarrow}$ has been suitably transformed into
external potentials. Adding these to the Kohn-Sham Hamiltonian led
to the effective Hamiltonian of Eq. (\ref{effhamilt}). As proven by
the Hohenberg-Kohn theorem, the electron density resulting from
occupying all single-particle states up to the common Fermi level
corresponds therefore to the fully relaxed electronic structure
under the given constraint. Calculating the total energy,
$E^e_{N_{\rm A}^{\uparrow}, N_{\rm A}^{\downarrow}, N_{\rm
B}^{\uparrow}, N_{\rm B}^{\downarrow}}$, connected to this electron
density provides then (together with $V^{\rm nuc-nuc}$) the
non-adiabatic PES $V^{\rm na}_{N_{\rm A}^{\uparrow}, N_{\rm
A}^{\downarrow}, N_{\rm B}^{\uparrow}, N_{\rm B}^{\downarrow}}$,
representing the chosen spin and electron numbers in the two
subsystems. Additionally, the derivatives of the non-adiabatic PES
with respect to the atomic positions can be obtained analytically in
the standard way providing the forces acting on the atoms.

At first glance, it appears as if our LC-DFT formalism builds on the
self-consistent solutions to the Kohn-Sham Hamiltonian and then
requires two additional nested sc loops. However, for the latter
self-consistency under the imposed constraint there is no need to
start with self-consistent Kohn-Sham solutions. Additionally, the
mere eigenvalue shift induced by the spin Fermi level alignment step
is nicely compatible with the structure of existing DFT codes. From
a computational point of view, our algorithm can thus be implemented
as one additional sc cycle at each iteration of the existing
electronic sc cycle in a DFT code. Of course, all known approaches
to improve the convergence of sc cycles like the use of
sophisticated mixing schemes can equally be applied to the
additional cycle. Having implemented a Pulay mixing scheme
\cite{pulay80}, our experience was that for the tested systems only
the very first DFT iterations required a significant number of inner
iterations, typically about 5 to 10, while close to the outer
self-consistency also the inner self-consistency was often directly
reached after the first iteration. For the example involving the
Al(111) surface, we even found the overall DFT convergence, i.e. the
number of iterations in the outer sc cycle, frequently much improved
compared to standard adiabatic calculations, because the oscillation
of states around the Fermi level is reduced when controlling the
electron numbers and thereby also the occupation numbers of the
Kohn-Sham states close to the Fermi level. For the systems with
localized electronic states, we found that using the spins or
electron numbers in the subsystems as convergence criterion for the
self-consistency was sometimes preferred to the equality of the
Fermi energies. Overall, for the systems studied, the cost of the
calculations using the constraint was thus often about the same and
sometimes even lower than the cost of standard adiabatic
calculations.

\subsection{Comparison to $\Delta$SCF and Fixed-Spin Moment}

An important characteristic of the LC-DFT approach presented here is
that it is parameter-free and only the set of the electron numbers
in the four channels defining the non-adiabatic state of interest
has to be specified. Under this constraint, the electronic structure
is fully relaxed, i.e. the partial densities of states are not
frozen and shifted statically. Instead, the single-particle states
are flexible to vary the contribution of each basis function to each
single-particle state freely. This can lead to a significant
improvement compared to the two prominent and widely employed
implementations of the constrained DFT concept, the $\Delta$SCF
\cite{gunnarsson76} and the fixed-spin moment approach
\cite{schwarz84}. In the latter approach the system is only
separated into a spin-up and a spin-down channel, which are filled
independently. In LC-DFT we go a step further and allow also to
distribute the spin-up and spin-down electrons in a well-defined way
into the two subsystems, which permits a more general control over
the spatial distribution of the electron and magnetization
densities. The different results obtained with both methods are
particularly obvious for the oxygen dissociation at Al(111) case
described in Section~IIIC below.

The improvement compared to the $\Delta$SCF method concerns extended
systems. Both approaches have in common that they consider two
subsystems and first analyze the pDOSs to determine the
contributions of each subsystem to the individual single-particle
states. However, in the $\Delta$SCF method the states with high
A-parts are then completely assigned to subsystem A, regardless of
their B-contributions, while the states with small A-contributions
are completely assigned to subsystem B.\cite{hellman04} A subsequent
occupation of the states by 0 or 1 electrons is therefore only fully
justified for the typically not interesting case of non-interacting
subsystems, i.e. no hybridization. Otherwise it results in
fractional effective occupation numbers of the individual
subsystems, which necessarily introduces some uncertainty in the
total energies obtained in the $\Delta$SCF method. The LC-DFT
approach, on the other hand, allows for a physical rehybridization
of the states {\em under the imposed constraint}. It thus conserves
the electron numbers of both subsystems, while at the same time
fully taking hybridization into account. This is the main difference
between the present LC-DFT formalism and the $\Delta$SCF method,
while in the limit of infinitely separated subsystems both
approaches are equivalent.

It is finally also important to note that all methods discussed
here, the FSM approach, the $\Delta$SCF method and the LC-DFT
formalism, intend to overcome limitations in the description of
chemical processes by controlling the electron numbers. This does
obviously not allow to overcome approximations in the employed
exchange-correlation (xc) functional. Local-density or
gradient-corrected xc functionals are e.g. known to cause
inaccuracies in the energy splittings between different spin
multiplets \cite{ziegler77,vonbarth79,jones89}. To this end, we note
that in the limit of infinite separation between the subsystems
(reactants), our approach reduces to the description of two isolated
subsystems. The non-adiabatic states of defined spin or charge in
our approach are then simply the corresponding excited states of the
isolated subsystems, i.e. of the {\em non-interacting} reactants. It
is worthwhile to point out that this is different to the diabaticity
concept, which reduces in this limit of infinite separation to the
adiabatic (excited) states of the {\em interacting} system. If there
is a (unphysical) charge or spin transfer even at these distances
(as in the example of O$_2$/Al(111) discussed below), it will be
present in both the adiabatic and the diabatic description, but not
in our non-adiabatic approach. The reduction to the states of the
non-interacting reactants provides also a possibility to check the
accuracy and suitability of the employed approximate xc functional
to describe the system states of interest. For the non-interacting
reactants it is typically feasible to check the results obtained
with the employed xc functional against higher-level calculations.
If one finds a sufficiently accurate description there, e.g. for the
lowest state of different symmetries, then we believe it is a
valuable approach to maintain this functional also for closer
distances of the reactants and get approximate insight into the
corresponding non-adiabatic potential-energy surface.

\section{Applications}

As an important application of the LC-DFT method we now illustrate
its use in the calculation of non-adiabatic PESs that are of
interest in the investigation of dynamic processes. Specifically, we
focus here on the scattering of atoms or molecules in crossed
molecular beams or at solid surfaces. As a side effect this also
shows how the method can be employed to suitably restrict unwanted
electron transfer between weakly or non-interacting subsystems. In a
DFT calculation, such an electron transfer will e.g. occur whenever
an occupied level of subsystem A is higher in energy than an
unoccupied level of subsystem B. Alignment of the Fermi levels of
the two systems will then lead to a fractional occupation of both
states in the self-consistent solution, even if the distance between
the two subsystems is macroscopic. A prominent example for this is
the interaction of an oxygen molecule with the Al(111) surface,
where the unoccupied $2\pi^{\ast\downarrow}$ orbitals of the
molecule are lower than the Fermi level of the
metal.\cite{yourdshahyan02,behler05} The resulting electron transfer
lowers the total energy of the system and at macroscopic distances
the latter does then not converge to the physically correct limit,
given by the sum of the total energies of the isolated molecule and
the isolated surface.

As an example for an extended periodic system, we correspondingly
briefly discuss the interaction of O$_2$ with the Al(111) surface.
In addition, we also present calculated non-adiabatic PESs for two
finite model systems, namely for the scattering of Na and Cl, and
the scattering of Cl$_2$ at a small Mg$_4$ cluster. All calculations
have been carried out using the all-electron DFT code DMol$^3$,
which employs numerical atomic-like orbitals as basis
functions.~\cite{delley90} Unless otherwise noted, we employ the
DMol$^3$ standard basis set ``all'' consisting of atomic orbitals
and polarization functions \cite{delley90}, a real-space cutoff of
12 bohr for the basis functions, and the PBE~\cite{perdew96} xc
functional.

\subsection{Na + Cl}

\begin{table}
\begin{ruledtabular}
\begin{tabular}{lrrrr}
\multicolumn{5}{c}{Na + Cl} \\ \hline
& $N_{\rm Na}^{\uparrow}$ & $N_{\rm Na}^{\downarrow}$
& $N_{\rm Cl}^{\uparrow}$ & $N_{\rm Cl}^{\downarrow}$ \\ \hline
Ionic            & 5 & 5 & 9 & 9 \\
Neutral, singlet & 6 & 5 & 8 & 9 \\
Neutral, triplet & 6 & 5 & 9 & 8 \\ \hline
\multicolumn{5}{c}{Mg$_4$ + Cl$_2$} \\ \hline
& $N_{\rm Mg_4}^{\uparrow}$ & $N_{\rm Mg_4}^{\downarrow}$
& $N_{\rm Cl_2}^{\uparrow}$ & $N_{\rm Cl_2}^{\downarrow}$ \\ \hline
Neutral        & 24 & 24 & 17 & 17 \\
Ionic, singlet & 23 & 24 & 18 & 17 \\
Ionic, triplet & 24 & 23 & 18 & 17 \\ \hline
\multicolumn{5}{c}{O$_2$ + Al(111)} \\ \hline
& $N_{\rm O_2}^{\uparrow}$ & $N_{\rm O_2}^{\downarrow}$
& $N_{\rm Al(111)}^{\uparrow}$ & $N_{\rm Al(111)}^{\downarrow}$ \\ \hline
Neutral, triplet & 18 & 14 & 409.5 & 409.5 \\ \hline
\end{tabular}
\end{ruledtabular}
\caption{Constrained electron numbers for the three test systems discussed.}
\label{tableI}
\end{table}

\begin{figure}[t!]
\scalebox{0.30}{\includegraphics{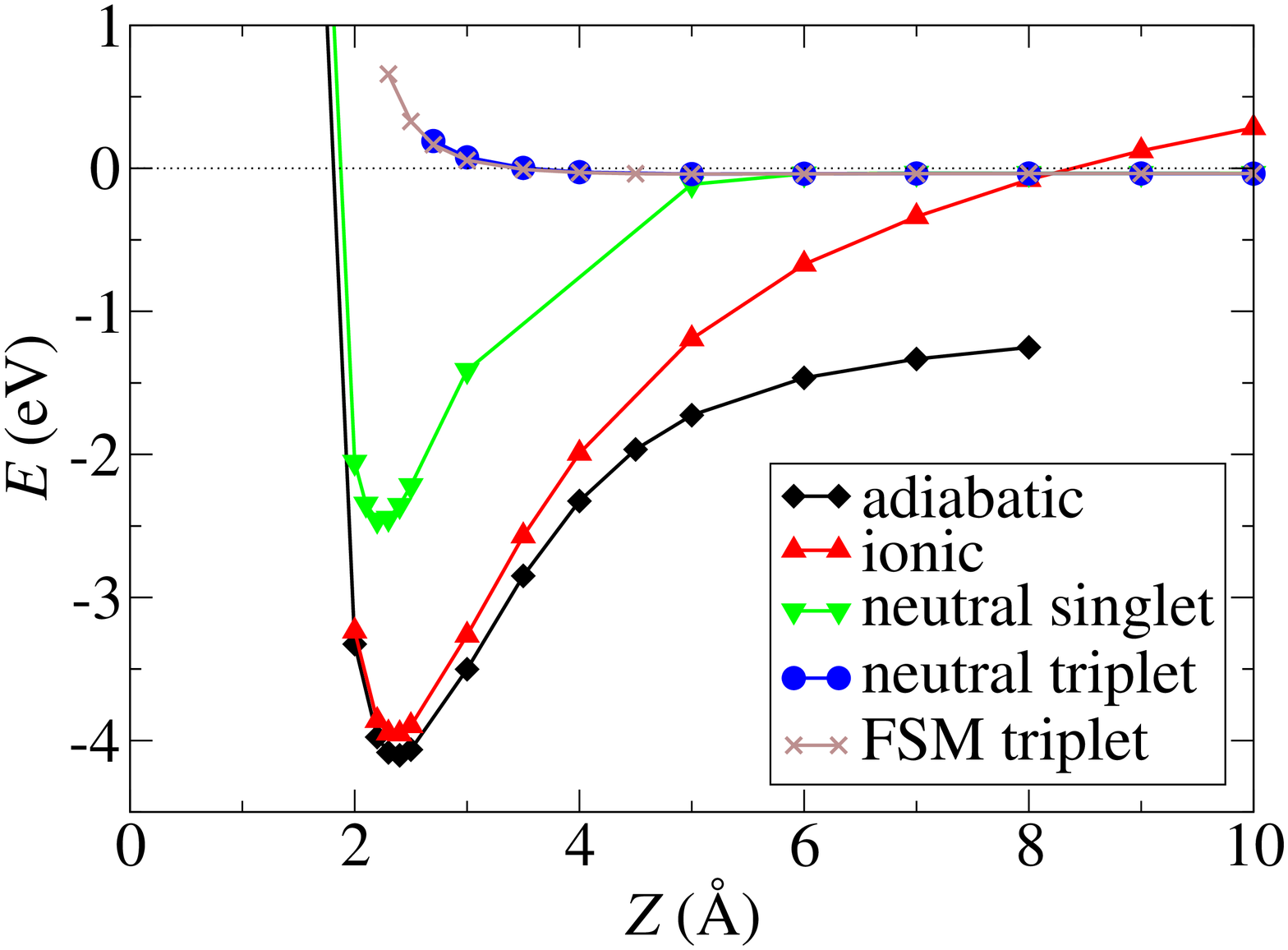}} \caption{(Color online)
Non-adiabatic potential-energy surfaces (PESs) for the scattering of a Na
and a Cl atom. Shown are the energies as a function of the
interatomic distance $Z$. See Table \ref{tableI} and text for the
constrained electron numbers defining the various non-adiabatic PESs.
Additionally shown are the adiabatic ground state PES and the PES
obtained with a fixed-spin moment (FSM) triplet calculation. The
energy zero corresponds to the energy of the two isolated, neutral
atoms.} \label{fig2}
\end{figure}

In the scattering of a sodium and a chlorine atom, two non-adiabatic
states of interest are the ionic PES ``Na$^+$ + Cl$^-$'' and the
neutral PES ``Na + Cl''. Since both neutral atoms are spin doublets
in their ground states, there are two possible relative orientations
of the spins in the latter case, yielding an overall singlet
(antiparallel spins) or triplet (parallel spins) neutral state.
Identifying each atom as one subsystem in our LC-DFT approach, Table
\ref{tableI} shows the constrained electron and spin numbers we used
to represent each of these non-adiabatic states. The resulting PESs
as a function of the interatomic separation are shown in
Fig.~\ref{fig2}, in which we additionally include the computed
adiabatic ground state PES and the PES as resulting from a FSM
calculation for an overall spin-triplet state ($N^{\uparrow}=15$,
$N^{\downarrow}=13$). In both the latter PES and the neutral triplet
PES there are therefore two unpaired electrons, but only the neutral
triplet PES computed by LC-DFT has the additional control of
locating one unpaired electron explicitly at each atom.

For bond lengths lower than 8~{\AA}, the energetically most
favorable non-adiabatic state is found to be the ionic PES, whereas
for larger distances these are the degenerate singlet and triplet
neutral PESs. The degeneracy of the latter two PESs is only lifted
for distances smaller than $~ 5$\,{\AA}, at which the Pauli
repulsion between the two unpaired spin-up electrons leads to a
strong increase of the neutral triplet PES. Interestingly, the FSM
curve is for all distances virtually degenerate to this neutral
triplet PES, indicating that even without constraint one unpaired
electron wants to stay at each atom. The minimum of the adiabatic
PES, on the other hand, is somewhat lower than the minimum of the
ionic PES, showing that the electron transfer in the adiabatic case
differs slightly from the one electron imposed in the LC-DFT
computation.

\begin{figure}[t!]
\scalebox{0.45}{\includegraphics{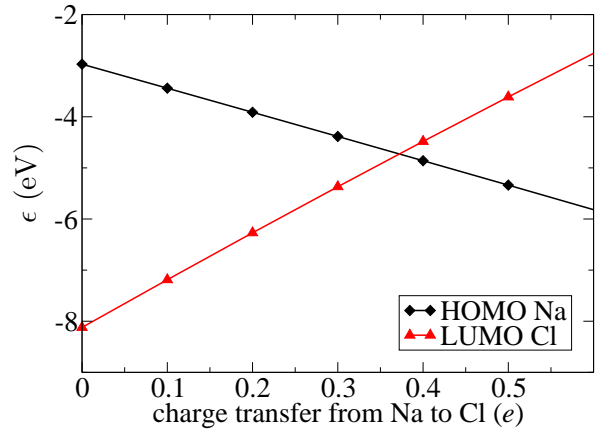}} \caption{(Color online)
Level energy of the highest occupied Kohn-Sham molecular orbital
(HOMO) of a free Na atom and of the lowest unoccupied Kohn-Sham
molecular orbital (LUMO) of a Cl atom as a function of the
electronic occupation. The occupations in the two separate
calculations of the isolated atoms are varied in form of a charge
transfer, i.e. the Na atom is computed with a fraction of an
electron removed, and the Cl atom is computed with this fraction of
an electron added.} \label{fig3}
\end{figure}

Another prominent feature of Fig.~\ref{fig2} is that for large
interatomic separations the adiabatic PES is about 1~eV lower than
the limit of neutral separated atoms. This is an illustration of the
above described electron transfer problem in adiabatic calculations.
Even for infinite interatomic distances, a small amount of charge is
transferred from the sodium to the chlorine atom, since the lowest
unoccupied 3$p^{\downarrow}$ state (LUMO) of the latter, is lower in
energy than the highest occupied 3$s^{\uparrow}$ state (HOMO) of the
sodium atom. In the self-consistent calculation, electron density is
consequently transferred, until the Fermi levels of the two atoms
are aligned. At infinite separation between the two atoms, this
charge transfer can be determined quantitatively by calculating the
Na Kohn-Sham HOMO and Cl Kohn-Sham LUMO level energies as a function
of different occupations. The results obtained from calculations of
the isolated charged atoms are displayed in Fig.~\ref{fig3} and show
that HOMO and LUMO are only aligned after an electron transfer of
0.37~$e$. This unphysical electron transfer at infinite separations
is not possible in the LC-DFT approach by construction, explaining
why in contrast to the adiabatic PESs the neutral triplet and
singlet PESs approach the correct limit.

\begin{figure}[t!]
\scalebox{0.3}{\includegraphics{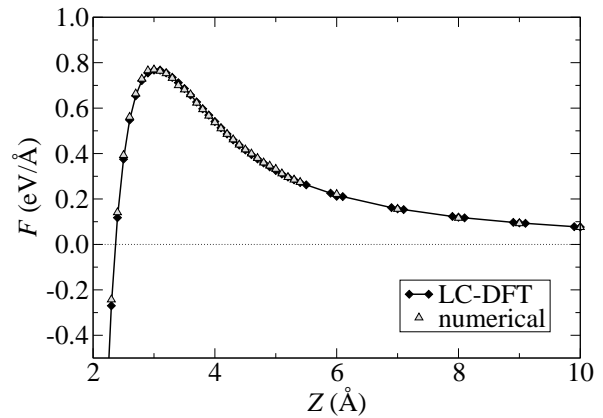}} \caption{Force acting on the
Cl atom in the NaCl dimer as a function of the interatomic distance
$Z$ for the ionic PES. Filled squares represent the forces
calculated within the LC-DFT approach, open triangles the forces
resulting from a numerical differentiation of the PES shown in Fig.
\ref{fig2}.} \label{fig4}
\end{figure}

Finally, we also employed this system to test the proper evaluation
of the forces in the constrained LC-DFT calculations. Taking the
example of the ionic PES, Fig.~\ref{fig4} shows the force on the Cl
atom computed either analytically within the LC-DFT approach or
numerically by differentiating the PES shown in Fig. \ref{fig2}. The
agreement is excellent proving that forces can be computed
accurately within the LC-DFT approach.

\subsection{Mg$_4$ + Cl$_2$}

\begin{figure}[t!]
\scalebox{0.5}{\includegraphics{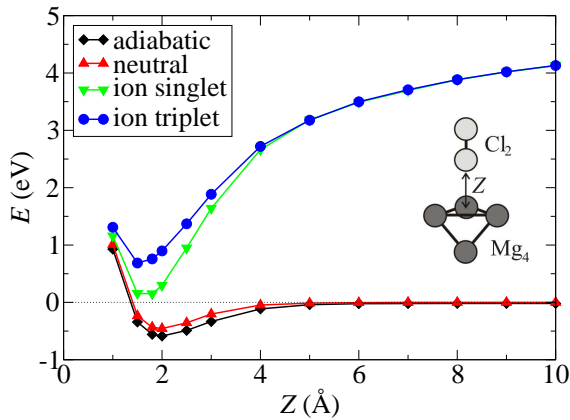}} \caption{(Color online)
Non-adiabatic potential-energy surfaces (PESs) of a Cl$_2$ molecule
scattering at a Mg$_4$ cluster. Shown are the energies as a function
of the distance $Z$ of the centers-of-mass of the Cl$_2$ molecule
and the Mg$_4$ cluster, with a scattering geometry as explained in
the inset. See Table \ref{tableI} and text for the constrained
electron numbers defining the various non-adiabatic PESs. Additionally
shown is the adiabatic ground state PES, and the energy zero
corresponds to the energy of the isolated neutral Cl$_2$ molecule
and Mg$_4$ cluster.} \label{fig5}
\end{figure}

As a simple example for subsystems consisting of groups of atoms we
discuss the interaction of a Cl$_2$ molecule with a small metal
cluster formed of a tetrahedron of 4 magnesium atoms. To illustrate
the method we restrict ourselves here to computing the PESs as a
function of the distance of a Cl$_2$ molecule approaching the Mg$_3$
plane of the cluster as explained in Fig. \ref{fig5}. For this we
first relaxed the structure of the cluster and the Cl$_2$ molecule
separately, and then held the resulting Mg-Mg bond lengths of
3.07\,{\AA} and the Cl-Cl bond length of 2.03\,{\AA} fixed in the
subsequent calculations. Using the electron numbers compiled in
Table \ref{tableI}, we calculated the non-adiabatic PESs
corresponding to the neutral, the ionic-singlet, and the
ionic-triplet state. Together with the adiabatic PES, the resulting
curves are shown in Fig. \ref{fig5}. At all distances, the
non-adiabatic state corresponding to a neutral configuration
exhibits the lowest energy, with the two ionic curves exhibiting
significantly higher energies. Similar to the Na+Cl case the latter
two become degenerate at larger distances, when the Pauli repulsion
affecting the ionic triplet curve becomes negligible. The closeness
of the non-adiabatic neutral curve to the adiabatic result indicates
only a comparably small electron transfer from the cluster to the
molecule during the scattering process. In this case, there is
therefore also only a small electron transfer problem and the
adiabatic curve approaches the proper limit for large
molecule-cluster separations. By differently occupying the Kohn-Sham
HOMO and LUMO levels as done in Fig. \ref{fig3}, we indeed obtain
only a very small electron transfer of 0.02~$e$ that is required to
align the Fermi energies in this case.

\subsection{O$_2$ Dissociation at Al(111)}

As a final example for an extended system, treated by periodic
boundary conditions, we turn to the dissociation of an O$_2$
molecule at the Al(111) surface. For this system the postulated
dominant role of non-adiabatic effects
\cite{yourdshahyan02,behler05,honkala00} could not be verified until
recently, since only empirical estimates of the underlying
non-adiabatic PESs were available
\cite{hellman03,kosloff04,hellman05}. Using LC-DFT we now focus on
the non-adiabatic neutral triplet PES, which is a suitable
representation at large distances from the surface, where the
gas-phase O$_2$ molecule will be in its spin-triplet ground state
and the Al(111) surface in a spin-singlet state. For the
calculations we employed a (3$\times$3) Al(111) slab consisting of 7
aluminium layers, separated by a 30~{\AA} vacuum. Oxygen is adsorbed
at both sides of the slab to establish inversion symmetry, a real
space cutoff of 9 bohr has been applied to the basis functions, and
10 $\mathbf{k}$-points have been used to sample the irreducible
wedge of the Brillouin zone. The electron numbers of the oxygen and
the aluminium subsystem used to define the neutral triplet PES are
listed in Table~\ref{tableI}.

\begin{figure*}
\scalebox{1.00}{\includegraphics{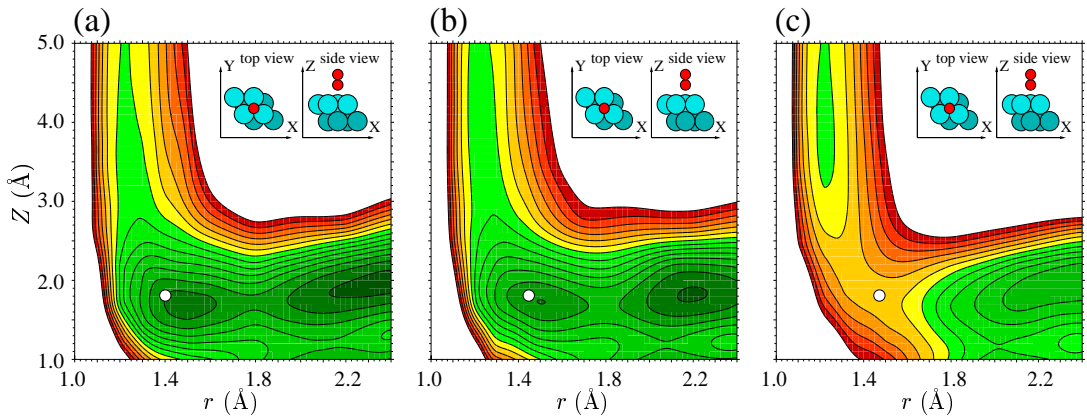}} \caption{(Color online)
Two-dimensional cut (``elbow plot'') through the high-dimensional
PES for the O$_2$ dissociation at Al(111). The energy is shown as a
function of the center-of-mass distance of the molecule from the
surface $Z$ and the oxygen-oxygen bond length $r$. The molecule
approaches the surface head-on above an fcc site as explained in the
insets. (a) Adiabatic calculation, (b) triplet fixed-spin moment
calculation, (c) neutral triplet LC-DFT calculation. Only the latter
PES exhibits an energy barrier. The energy difference between the
contour lines is 0.2~eV and the small white circle denotes the
molecular position discussed in Fig. 7.} \label{fig6}
\end{figure*}

\begin{figure}[t!]
\scalebox{0.5}{\includegraphics{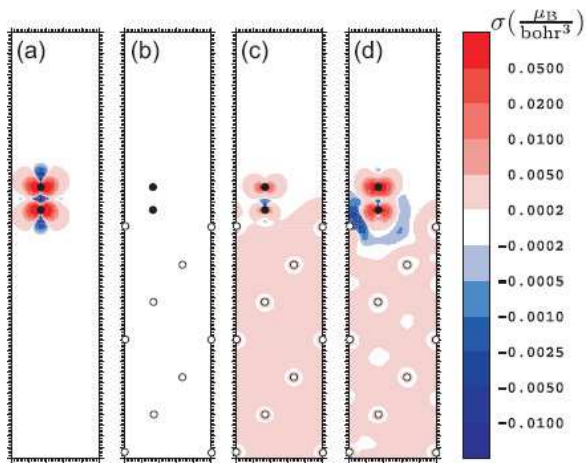}} \caption{(Color online)
Magnetization density (difference between spin-up and spin-down
electron density) for a molecule at the energy barrier of the LC-DFT
triplet PES ($r = 1.4$\,{\AA}, $Z = 1.8$\,{\AA}, as marked in Fig.
\ref{fig6}). Shown is a two-dimensional cut perpendicular to the
surface and through the O$_2$ molecule, along the $[01\bar{1}]$ and
$[111]$ direction. The position of the two O atoms are marked as
small black circles, the position of the Al atoms as small white
circles. (a) Isolated O$_2$ molecule in its spin-triplet ground
state, (b) adiabatic calculation for O$_2$/Al(111), (c) triplet FSM
calculation for O$_2$/Al(111), and (d) neutral triplet LC-DFT
calculation for O$_2$/Al(111).} \label{fig7}
\end{figure}

Discussing our results for the high-dimensional PES in detail
elsewhere \cite{behler05,behler06}, we illustrate the insights
gained by the LC-DFT approach by concentrating on the
two-dimensional dependence on the molecular bond length $r$ and the
center-of-mass distance of the molecule from the surface $Z$ for a
fixed molecular orientation and lateral position over the surface.
Fig.~\ref{fig6} shows corresponding ``elbow plots'' specifically for
an O$_2$ molecule approaching the surface head-on and above an fcc
site. In agreement with previous studies
\cite{yourdshahyan02,honkala00} the adiabatic PES displayed in
Fig.~\ref{fig6}a does not exhibit an energy barrier to dissociation,
a finding that cannot be reconciled with the experimentally
well-established low sticking coefficient for thermal
molecules~\cite{behler05,osterlund97}. Suspecting a dominant role of
non-adiabatic effects as the reason for this discrepancy, we turn to
non-adiabatic representations, in which particularly the spin-triplet
character of the gas-phase O$_2$ molecule is conserved.
Fig.~\ref{fig6}b and c show corresponding PESs obtained with the FSM
approach for an overall triplet state of the system and with our
LC-DFT approach constraining the spin-triplet to the O$_2$ molecule,
respectively. In contrast to the Na+Cl neutral triplet PES discussed
above, the FSM and LC-DFT approach now yield qualitatively different
results. While no barrier is obtained in the prior method, the
neutral triplet PES calculated with LC-DFT exhibits a clear energy
barrier.

The reason for this difference is the different distribution of the
magnetization density in the system. This is illustrated for a
molecular configuration at the energy barrier in Fig. \ref{fig7}. In
(a) the magnetization density computed for the free oxygen molecule
(i.e. without Al(111) slab) in its spin-triplet ground state is
shown, whereas (b) displays the result of an adiabatic calculation
including the Al(111) slab. In the latter case, neither the O$_2$
molecule, nor the metal atoms exhibit any spin-polarization, which
is the most favored state for small molecule-surface separations. In
the FSM calculation, the total spin of the system is forced to be a
triplet, but as apparent from (c) the majority of this excess spin
is not located at the O$_2$ molecule, but distributed over the
entire metal slab. In contrast to this, in the LC-DFT result shown
in (d) the triplet spin is localized at the oxygen molecule,
reflecting the improved control over the spatial distribution of the
magnetization density in the latter approach. The accumulation of
spin-up density on the O$_2$ molecule repels the spin-up density of
the metal slab towards the interior of the slab, while there is a
strong accumulation of spin-down density close to the molecule. As a
consequence, the metal slab is still in an overall singlet state in
the LC-DFT calculation, which is obviously a better representation
of the non-adiabatic state defined by an impinging triplet O$_2$
molecule compared to the FSM result. In addition, the LC-DFT
approach overcomes the small charge transfer problem present in this
system, as well. Since the unoccupied $2\pi^{\ast\downarrow}$
orbitals of the O$_2$ molecule are lower than the Fermi level of the
metal, the adiabatic calculation yields a charge transfer of
0.01~$e$ to the O$_2$ molecule even at macroscopic distances from
the surface.

\section{Summary}

In summary, we presented a locally-constrained density-functional
theory (LC-DFT) approach, that allows to confine electrons to
sub-spaces of the Hilbert space, e.g. to selected atoms or groups of
atoms. A major application of this technique is the computation of
non-adiabatic potential-energy surfaces, which we illustrated with
examples treating the scattering of atoms and molecules at other
atoms, clusters or surfaces. Following the general formulation by
Dederichs {\em et al.} \cite{dederichs84}, the electron confinement
is realized by suitably introducing additional constraints to the
electronic Hamiltonian. DFT is then used to obtain the fully relaxed
electronic structure under the additional external potential imposed
by the applied constraints. With the $\Delta$SCF and FSM methods as
widely employed alternative implementations of this general concept,
our LC-DFT method offers a more systematic approach to extended
systems compared to $\Delta$SCF, and better control over the spatial
distribution of the constraint electrons compared to FSM. This
better spatial control allows also to overcome the charge transfer
problem between widely separated subsystems that can occur in
adiabatic DFT calculations.

\section{Acknowledgements}

We thank Volker Blum for carefully reading the manuscript and useful
suggestions. Stimulating discussions with Bengt I. Lundqvist and
Eckart Hasselbrink are gratefully acknowledged.

\section*{Appendix: The Projection Operator}

In general, localized atom-centered basis functions like atomic
orbitals are not orthogonal, and a projection operator should be
formulated in terms of covariant and contravariant basis
functions~\cite{headgordon98,artacho91}. The covariant Bloch basis
functions $|\chi_{i}^{\mathbf{k}}>$ and the contravariant Bloch
basis functions $|\chi^{j \mathbf{k}}>$ are related to each other by
the equations
\begin{subequations}
\label{cocontra}
\begin{eqnarray}
<\chi^{j \mathbf{k}}| &=& \sum_{i=1}^{n} \;
(S^{\mathbf{k}}_{ij})^{-1} <\chi_{i}^{\mathbf{k}}| \quad,\\
|\chi^{j \mathbf{k}}> &=& \sum_{i=1}^{n} \;
|\chi_{i}^{\mathbf{k}}> (S^{\mathbf{k}}_{ij})^{-1} \quad,\\
< \chi_i^{\mathbf{k}}| &=& \sum_{j=1}^{n} \;
S_{ji}^{\mathbf{k}} <\chi^{j\mathbf{k}}| \quad,\\
{\rm and} \quad
|\chi_i^{\mathbf{k}}>  &=& \sum_{j=1}^{n} \;
|\chi^{j \mathbf{k}}>  S_{ji}^{\mathbf{k}} \quad,
\end{eqnarray}
\end{subequations}
with $\mathbf{S}^{\mathbf{k}}$ being the overlap matrix. For
covariant and contravariant basis functions the following
orthonormality relation holds,
\begin{equation}
<\chi^{i\mathbf{k}}|\chi_j^{\mathbf{k}}> \;=\;
<\chi_i^{\mathbf{k}}|\chi^{j\mathbf{k}}> \;=\; \delta_{ij} \quad,
\end{equation}
where $\delta_{ij}$ is the Kronecker symbol.

In principle there are two possible forms of the projection operator
$\mathbf{P}_{\rm A}^{\mathbf{k}}$ into the subsystem A, which are
equivalent,
\begin{subequations}
\begin{eqnarray}
\mathbf{P}_{\rm A}^{\mathbf{k}} &=& \sum_{i=1}^m
|\chi_i^{\mathbf{k}}><\chi^{i\mathbf{k}}|  \label{proj1}\\
\mathbf{P}_{\rm A}^{\mathbf{k}} &=& \sum_{i=1}^m
|\chi^{i\mathbf{k}}><\chi_i^{\mathbf{k}}| \quad , \label{proj2}
\end{eqnarray}
\end{subequations}
and where the sum $i=1\ldots m$ runs over the $m$ basis functions of
subsystem A. However, expanded onto the Bloch basis functions, both
these formulations for $\mathbf{P}_{\rm A}^{\mathbf{k}}$ yield
non-hermitian matrices. In order to facilitate the implementation
into existing DFT codes, we therefore prefer to work with the
following symmetrized form, which does lead to a hermitian matrix
\begin{equation}
\mathbf{P}_{\rm A}^{\mathbf{k}} \;=\;
\frac{1}{2} \cdot \left( \sum_{i=1}^m|\chi_{i}^{\mathbf{k}}><\chi^{i\mathbf{k}}| \;+\; \sum_{i=1}^m |\chi^{i\mathbf{k}}><\chi_{i}^{\mathbf{k}}| \right) \quad .
\end{equation}

Inserting the expressions for the contravariant basis functions in
Eq.~(\ref{cocontra}) enables us to express the projection operator
entirely in terms of the known covariant basis functions (the atomic
orbitals)
\begin{eqnarray}
\mathbf{P}_{\rm A}^{\mathbf{k}} &=&
\frac{1}{2} \cdot \left( \sum_{i=1}^m \sum_{j=1}^n
|\chi_{i}^{\mathbf{k}}> (S^{\mathbf{k}}_{ji})^{-1}
<\chi_{j}^{\mathbf{k}} | \right. \\ \nonumber
&& \left. +\; \sum_{i=1}^m \sum_{j=1}^n
|\chi_{j}^{\mathbf{k}}> (S^{\mathbf{k}}_{ji})^{-1}
<\chi_{i}^{\mathbf{k}} | \right) \quad .
\end{eqnarray}
Starting from this expression, we can then derive the matrix
representation of this operator in the Hilbert space spanned by the
Bloch basis functions
\begin{eqnarray}
P_{\rm A,ij}^{\mathbf{k}} &=&
<\chi_i^{\mathbf{k}}| \mathbf{P}_{\rm A}^{\mathbf{k}} |\chi_j^{\mathbf{k}}> \nonumber \\
&=& \frac{1}{2} <\chi_i^{\mathbf{k}}|
\left( \sum_{k=1}^m \sum_{l=1}^n
|\chi_{k}^{\mathbf{k}}> (S^{\mathbf{k}}_{lk})^{-1}
<\chi_{l}^{\mathbf{k}}| \right. \nonumber \\
& & \left. +\; \sum_{r=1}^m \sum_{s=1}^n
|\chi_{s}^{\mathbf{k}}> (S^{\mathbf{k}}_{sr})^{-1}
<\chi_{r}^{\mathbf{k}}| \right) |\chi_j^{\mathbf{k}}> \nonumber\\
&=& \frac{1}{2} \left( \sum_{k=1}^m \sum_{l=1}^n <\chi_i^{\mathbf{k}}|\chi_{k}^{\mathbf{k}}>
(S_{lk}^{\mathbf{k}})^{-1}
<\chi_{l}^{\mathbf{k}}|\chi_{j}^{\mathbf{k}}> \right. \nonumber\\
& & \left. +\; \sum_{r=1}^m \sum_{s^=1}^n <\chi_{i}^{\mathbf{k}}|\chi_{s}^{\mathbf{k}}>
(S_{sr}^{\mathbf{k}})^{-1}
<\chi_{r}^{\mathbf{k}}|\chi_{j}^{\mathbf{k}}> \right) \nonumber\\
&=& \frac{1}{2} \left( \sum_{k=1}^m \sum_{l=1}^n
S_{ik}^{\mathbf{k}} (S_{lk}^{\mathbf{k}})^{-1} S_{lj}^{\mathbf{k}}
\right. \nonumber \\
& & \left. +\; \sum_{r=1}^m \sum_{s=1}^n
S_{is}^{\mathbf{k}} (S_{sr}^{\mathbf{k}})^{-1} S_{rj}^{\mathbf{k}}
\right) \nonumber\\
&=& \frac{1}{2} \left( \sum_{k=1}^m S_{ik}^{\mathbf{k}} \delta_{kj} \;+\; \sum_{r=1}^m \delta_{ir} S_{rj}^{\mathbf{k}} \right) \nonumber\\
&=& \Bigg\{
\begin{array}{rcl}
S_{ij}^{\mathbf{k}} & \mbox{for} & i\leq m \;\wedge\; j\leq m \\
\frac{1}{2}S_{ij}^{\mathbf{k}}& \mbox{for}  & i\leq m \;\vee\; j \leq m\\ 0& \mbox{for}  & i>m \;\wedge\; j>m\quad .
\end{array}
\end{eqnarray}

\end{document}